\documentclass[sigconf,nonacm]{acmart}
\usepackage{amsmath,amssymb,booktabs,array}
\usepackage{algorithm}
\usepackage{algpseudocode}
\usepackage{tikz}
\usetikzlibrary{arrows.meta,positioning,fit,backgrounds}
\usepackage{microtype}
\newcommand{\miner}{\textsc{PSIMiner}}
\newcommand{\obs}{\mathcal{O}}
\newcommand{\rules}{\mathcal{R}}
\newcommand{\trace}{\sigma}

\usepackage{etoolbox}
\usepackage{balance}
\AtEndEnvironment{proof}{\vspace{-6pt}}
\AtEndEnvironment{theorem}{\vspace{-3pt}}
\begin{document}

\title{Mine Locally, Verify Globally: Exact Recovery, Compilation, and
       Distortion of Mined Temporal Rules}
\subtitle{Full Version}
\thanks{Full version of a short paper accepted at the 28th International Conference on Distributed Computing and Networking (ICDCN 2027), Surathkal, India, January 4--7, 2027. This is the authors' version, released under CC-BY 4.0.}
\renewcommand{\shorttitle}{Mine Locally, Verify Globally (Full Version)}

\author{Nikolaos Kekatos}
\affiliation{\institution{Clone Systems}\city{Larnaca}\country{Cyprus}}
\email{nkekatos@clone-systems.com}

\author{Priyanka Sinha}
\affiliation{\institution{Docyt}\city{Kolkata}\country{India}}
\email{priyanka.sinha.iitg@gmail.com}

\author{Stylianos Basagiannis}
\affiliation{\institution{International Hellenic University}\city{Serres}\country{Greece}}
\email{basagiannis@ihu.gr}

\author{Antonio Anastasio Bruto da Costa}
\affiliation{\institution{Pure Silicon Solutions}\city{Birmingham}\country{United Kingdom}}
\email{antoniobdacosta@gmail.com}

\author{Alexios Lekidis}
\affiliation{\institution{University of Thessaly}\city{Larissa}\country{Greece}}
\email{alekidis@uth.gr}

\author{Pabitra Mitra}
\affiliation{\institution{Indian Institute of Technology Kharagpur}\country{India}}
\email{pabitra@gmail.com}

\author{Tom Nianios}
\affiliation{\institution{Clone Systems}\city{Larnaca}\country{Cyprus}}
\email{tnianios@clone-systems.com}

\renewcommand{\shortauthors}{Kekatos et al.}

\begin{abstract}
Specification mining extracts temporal rules that describe relationships between events in
recorded system traces. Applying these rules in distributed monitoring systems is more
difficult because observations come from different nodes, whose clocks may not be
synchronized, and events may be lost or reordered during transmission. In this paper, we study how to aggregate the coverage of mined rules across observers, determine whether their temporal windows are needed during monitoring, and characterize their detection under event loss. Our evaluation uses IPv6 packet traces collected from $21$ receivers
across four continents. We show that local mining followed by global confirmation recovers centralized rule coverage
under suitable conditions while substantially reducing communication overhead. In contrast,
merging traces with unaligned clocks can distort coverage by introducing incorrect temporal
matches. A window-gain criterion identifies rules that require temporal state, while our loss
analysis and broker experiments demonstrate why contiguous losses can be more damaging than
independent losses. The results identify separate requirements for aggregating observations, selecting monitor state, and handling event loss.
\end{abstract}

\keywords{specification mining, runtime verification, distributed monitoring,
temporal logic, network measurement, fault tolerance}

\maketitle
\section{Introduction}\label{sec:intro}

Runtime verification checks a system while it runs. It is useful when correctness cannot
be settled before deployment~\cite{pnuelizaks,rvltl,whatcanyouverify}. A monitor needs
two things, a property to check and evidence to check it against, and each has its own
literature. Writing properties by hand scales poorly. Specification mining therefore
infers them from observed behaviour, as data invariants~\cite{daikon}, as temporal
rules~\cite{perracotta,javert,texada}, or as rules with explicit time
bounds~\cite{psiminer}. Evidence in a distributed system is spread over many nodes.
Decentralised runtime verification therefore splits a formula across the nodes that
observe it and combines their verdicts~\cite{sen2004,bauerfalcone,colombofalcone,themis}.
Layered monitoring architectures add delivery and continuity guarantees, so that a
verdict keeps its meaning when a component crashes or is overloaded~\cite{fabric}.

These two lines of work rarely meet. Miners run centrally and offline. A mined rule is
judged by how well it scores on the traces it was learned from. Monitors execute
properties that a person wrote, online, over a network that may drop, delay and reorder
events. Nobody has measured what a mined rule's verdict is worth when its evidence
arrives incomplete or out of order. Runtime verification does study imperfect
observation, but mostly as missing
events~\cite{rvse,adaptiverv,particlerv,samplingrv,lossytraces}. Disorder is treated as
a problem for the monitoring algorithm, which should undo it~\cite{outoforder}.

We met this question in a concrete setting. Our dataset comes from an active-probing
campaign. Twenty-one receivers on four continents record which IPv6 extension headers
survive transit~\cite{james}. The properties an operator would want have exactly the
shape of a mined rule, for example ``if the payload is not medium-sized, then no
fragment header follows within four seconds''. The monitoring fabric that would run
such a rule is also real, with bounded buffers, a crash window and a wide-area
path~\cite{fabric}. We expected loss to be the main problem, since lost events cost
detections. The experiments showed a different picture. Our rules ask for \emph{some}
matching event inside a window. Losing one candidate is cheap when the window holds many.
Reordering the stream, on the other hand, moves extra candidates into windows and
creates matches that never occurred. Merging traces from observers whose clocks all
start at zero does the same, because a window at one receiver can then be satisfied by
an event recorded on another continent. A monitoring system pays for false
alarms~\cite{baserate,closedworld}, and these invented matches become false alarms.

We organise the paper around three questions. Each answer has a proof and a measurement.

\emph{Where should mining run?} Coverage, the score we use, adds up across observers.
Mining at each observer and exchanging short summaries is therefore exact, not an
approximation (Sect.~\ref{sec:distmine}). The step that needs justification is the usual
one, central mining over merged traces.

\emph{Which mined rules need a temporal monitor?} Earlier work found that the main rule
of this dataset is a co-occurrence inside a single packet. Its time window does
nothing~\cite{csr}. We turn that observation into a \emph{compilation decision} that
applies to any rule (Sect.~\ref{sec:gate}).

\emph{What does the transport do to a deployed rule?} It distorts the score in both
directions. In the cases we measured, invented matches were the more frequent
direction (Sect.~\ref{sec:degrade}). Loss follows a law indexed by redundancy inside
each window. A law indexed by rule length gives wrong answers. Reordering is harder to
summarise. Very large displacement ends at the rule's same-event version, which we
define in Sect.~\ref{sec:gate}. The shuffle test of Sect.~\ref{sec:gate} measures
something else. Neither of the two bounds the effect of moderate displacement, which can
exceed both when the displacement scale is close to the rule's window.

\begin{figure*}[t]
\centering
\footnotesize
\resizebox{\textwidth}{!}{%
\begin{tikzpicture}[
  node distance=3.2mm,
  box/.style={draw, rounded corners=1.2pt, align=center, inner sep=2.2pt,
              minimum height=7.5mm, font=\footnotesize},
  obs/.style={box, fill=black!4, minimum width=13mm},
  stage/.style={box, fill=black!7, minimum width=21mm},
  res/.style={box, minimum width=28mm, align=left},
  note/.style={font=\scriptsize\itshape, align=center, text=black!65},
  ar/.style={-{Latex[length=1.6mm]}, thick},
]
\node[obs] (o1) {$\trace_1$};
\node[obs, below=1.2mm of o1] (o2) {$\trace_2$};
\node[obs, below=1.2mm of o2] (od) {$\vdots$};
\node[obs, below=1.2mm of od] (oN) {$\trace_N$};
\node[note, above=0.8mm of o1] {\textbf{21 observers}};
\node[stage, right=9mm of o2, yshift=-2mm] (mine) {Mine locally\\(Round 1)};
\draw[ar] (o1.east) -- (mine.west);
\draw[ar] (o2.east) -- (mine.west);
\draw[ar] (oN.east) -- (mine.west);
\node[stage, right=6mm of mine] (desc) {Send rule\\descriptions};
\draw[ar] (mine) -- (desc);
\node[note, below=0.6mm of desc] {$1658\times$ smaller\\than the traces};
\node[stage, right=6mm of desc] (merge) {Merge and confirm\\(Rounds 2--3)};
\draw[ar] (desc) -- (merge);
\node[note, below=0.6mm of merge] {traces must be separated\\in time; else $+79.9$\,pp};
\node[stage, right=6mm of merge, fill=black!12] (gate) {Gate\\window gain $W$};
\draw[ar] (merge) -- (gate);
\node[res, right=6mm of gate, yshift=8.5mm] (sl)
  {$W\le\delta_W$: \textbf{stateless} check\\
   no window, nothing to reorder};
\node[res, right=6mm of gate, yshift=-8.5mm] (wd)
  {$W>\delta_W$: \textbf{windowed} monitor\\
   ordering need decided by $D$};
\draw[ar] (gate.east) -- (sl.west);
\draw[ar] (gate.east) -- (wd.west);
\node[stage, right=6mm of sl.east, yshift=-8.5mm, minimum width=24mm] (dep)
  {Deploy on fabric\\tier and guarantees};
\draw[ar] (sl.east) -- (dep.west);
\draw[ar] (wd.east) -- (dep.west);
\node[note, below=0.6mm of dep, text width=26mm]
  {delivery guarantee priced\\from the loss law};
\end{tikzpicture}}
\caption{Overview of \textsc{Mine--Merge--Confirm--Gate} (Algorithm~\ref{alg:mmcg}) on
$21$ observers holding $757{,}932$ packets. No raw trace crosses the network. Observers
exchange rule descriptions, count pairs and scalar summaries. For the mining rounds
this costs $1658\times$ less than shipping the traces, and it is exact when the traces
are separated in time by more than a rule spans (Thm.~\ref{thm:recovery}). The gate uses
the window gain $W$ to decide whether a rule keeps window state (Thm.~\ref{thm:gate}).
The ordering guarantee is decided from $D$, the largest score change over the
displacements the deployment admits (Prop.~\ref{prop:refute}). The delivery guarantee
comes from the loss law (Thm.~\ref{thm:loss}).}
\label{fig:overview}
\end{figure*}
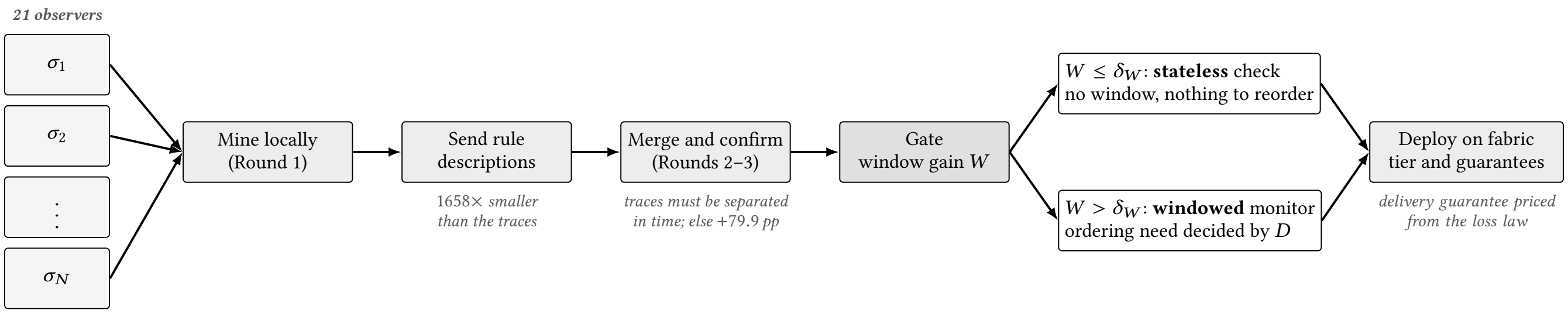

\paragraph{Contributions.}
Figure~\ref{fig:overview} shows the pipeline end to end.
\begin{itemize}
\item A \textbf{distributed mining protocol} with an exact recovery identity
(Thm.~\ref{thm:recovery}) and a communication bound (Prop.~\ref{prop:comm}). We also
measure the \emph{merging inflation}: $79.9$ percentage points on one rule, removed
completely when the observers' clocks are offset.
\item A \textbf{two-stage compiler pass} (Sect.~\ref{sec:gate}). The first stage tests
whether a rule needs its window and removes unneeded window state, with a soundness
result (Thm.~\ref{thm:gate}). The second stage measures how sensitive the remaining
windowed monitors are to event order.
\item A \textbf{detection law} under loss (Thm.~\ref{thm:loss}). An \textbf{exact
characterisation} of where large displacement ends, together with a proof that the
shuffle test does not reach that end point (Prop.~\ref{thm:inflate}). A
\textbf{counterexample} to the natural conjecture that the shuffle gap bounds the effect
of displacement (Prop.~\ref{prop:refute}).
\end{itemize}
The 5-page conference version presents the first result in full and the other two in
short form. This version adds all proofs, the reordering analysis, the false-alarm study
and the broker measurements.

The rest of the paper is organised as follows. Section~\ref{sec:setting} defines the
system, the rule model and the transport. Section~\ref{sec:distmine} gives the distributed
mining protocol, the theorem that it recovers central coverage exactly, and the measurements.
Section~\ref{sec:gate} introduces the compilation gate: the window gain, the shuffle gap
and the out-of-sample check. Section~\ref{sec:degrade} studies loss, reordering and the false
alarms they cause, and measures real brokers. Section~\ref{sec:threats} states the threats to
validity, Section~\ref{sec:related} reviews related work, and Section~\ref{sec:conclusion}
concludes. Appendix~\ref{app:proofs} contains the proofs.

\section{System and Rule Model}\label{sec:setting}

This section fixes three things: the shape of a mined rule, what an observer holds, and
which timestamps a monitor reads. It also defines the terms used in the rest of the
paper.

\begin{table}[t]
\centering\small
\caption{Notation. The upper block is fixed by the dataset and the rule language. The
lower block is computed or set by the deployment.}
\label{tab:notation}
\setlength{\tabcolsep}{3pt}
\begin{tabular}{@{}llll@{}}
\toprule
$\obs$, $N$ & observers, $N=|\obs|$ & $\varphi$, $k$ & rule \eqref{eq:template}; its window\\
$\trace_i$ & trace at observer $i$ & $\varphi^{(0)}$ & same-event version\\
\midrule
$\rules_i$ & rules mined at $i$ & $\tau_\ell$, $\tau_g$ & local, global thresholds\\
$c_i,s_i$ & local coverage; target count & $c^{\cup}$ & coverage of the union\\
$W$ & window gain \eqref{eq:wgain} & $\Delta$ & shuffle gap \eqref{eq:delta}\\
$D$ & largest score change over $R$ & $\delta_W,\delta_R$ & tolerances\\
$a$ & irreplaceable events & $q$, $m_\ell$ & windows; candidates in $\ell$\\
$p$, $r$ & loss rate; displacement width & $p^\ast$, $\eta$ & loss budget; req.\ detection\\
\bottomrule
\end{tabular}
\end{table}

\subsection{Rules}
Table~\ref{tab:notation} collects the notation. We use the bounded-past rule language of
\miner~\cite{psiminer}. A rule has the shape
\begin{equation}\label{eq:template}
B_n \; \#\#[0{:}k] \; \cdots \; \#\#[0{:}k] \; B_0 \;\longrightarrow\; E .
\end{equation}
Each \emph{bucket} $B_j$ is a set of Boolean conditions on the attributes of one event.
$E$ is the \emph{target} condition. The symbol $\#\#[0{:}k]$ means that the next bucket
holds at most $k$ time units later. A \emph{match} of the rule for a target event is a
choice of events, one per bucket, that satisfy the buckets and respect these time
bounds. We call such a choice a \emph{witness}.

The time bound is \emph{existential}. The rule asks that \emph{some} matching event lies
inside the window. It does not ask for a particular event. All results in
Sect.~\ref{sec:degrade} depend on this property.

The template leaves one convention open, and we fix it. Two different buckets \emph{may}
be satisfied by the same event when the time bounds allow it. This choice is used twice.
The same-event version of Sect.~\ref{sec:gate} relies on it. The loss law of
Thm.~\ref{thm:loss} needs witness events to be distinct, and states this as an explicit
assumption.

Each rule has two scores. \emph{Coverage} is the fraction of target events whose
window contains a matching antecedent. It equals $n^{\mathrm{expl}}/s$, where $s$ is
the number of target events and $n^{\mathrm{expl}}$ is the number of those that the rule
explains. \emph{Confidence} is the reverse conditional, the fraction of antecedent
matches followed by a target. We report the two separately, because mixing them makes
association metrics misleading~\cite{csr}. \textbf{The protocol of
Sect.~\ref{sec:distmine} thresholds on coverage only.} This is what makes
Thm.~\ref{thm:recovery} apply. Coverage is a ratio of counts, and counts add up over
observers. With a threshold on both metrics, the observer with the highest coverage and
the observer with the highest confidence can differ. Then no single observer needs to
propose the rule. Extending recovery to a two-metric acceptance rule is open.

\subsection{Observers}
An \emph{observer} is a node that holds its own trace and keeps its own clock. A
distributed dataset is a family of traces $\{\trace_i\}_{i\in\obs}$, one per observer.
Both properties are used below. Locality is what lets mining be distributed. Separate
clocks are what make naive merging unsound.

Our dataset is the JAMES IPv6 extension-header dataset~\cite{james}. It has $21$
controlled receivers on four continents (called vantage points in that work) and
$757{,}932$ packets. Each receiver captures what reaches it from a common set of
senders. Header stripping in transit is therefore observable. We project each trace to
a fixed per-packet schema that uses only IPv6 header fields.

Every receiver starts its clock at zero, and each trace spans roughly $4000$ to
$5400$\,s. Nothing in the dataset aligns the clocks. Section~\ref{sec:distmine} shows
what this costs when the traces are merged.

\paragraph{Key terms.}
Two terms recur and are defined here. \emph{Exact recovery} means that a distributed
procedure reports the same coverage as a central miner run on correctly separated traces.
It is an identity, not an approximation. \emph{Merging inflation} is the extra coverage
that appears when all traces are placed on one time axis. A window opened at one
receiver can then be satisfied by an event recorded at another.

\subsection{Transport}
Deployed monitors run over a monitoring fabric~\cite{fabric}. It has a message broker
that receives events from devices and a durable broker that forwards them from the
gateway to the back end. A small service attaches a timestamp to each event, and a local
store keeps outgoing messages until their delivery is confirmed. The fabric offers
guarantees on delivery, ordering, clock progress, isolation between consumers, and flow
control.

We evaluate rules over the timestamps \emph{seen at the monitoring tier}. These may be
arrival times or broker timestamps. Transport delay therefore changes the time values
that a rule uses. A deployment can avoid this by keeping trusted source timestamps and
re-sorting events by source time before evaluation. That costs buffering, clock
synchronisation and latency. This re-sorting is what the fabric's ordering guarantee
provides. The guarantees limit two remaining effects without removing them, and these
two effects are our parameters. The first is per-event loss with rate $p$. It comes from
bounded buffers under sustained overload and from the crash window between delivery
and durable storage. The second is reordering by up to $r$ time units on a wide-area
path.
\section{Distributed Specification Mining}\label{sec:distmine}

A dataset collected at many observation points is normally mined by gathering all traces
and mining their union. For coverage-based mining this step is unnecessary, because
coverage adds up over observers. It is also unsafe. The observers of a real campaign do
not share a clock, and merging their traces creates matches across observers. This
section gives the protocol, the identity that justifies it, and the measured cost of
naive merging.

\subsection{Protocol}
The protocol is called \textsc{Mine--Merge--Confirm} and runs in three rounds.
Algorithm~\ref{alg:mmcg} gives the full procedure, including the gate of
Sect.~\ref{sec:gate}.

\emph{Round 1 (mine).} Each observer $i$ runs the miner on its own trace with a local
threshold $\tau_\ell$ and obtains a set of candidate rules $\rules_i$. No trace leaves
the observer. For every candidate, the observer sends a \emph{rule description} to the
gateway. A description lists the buckets, the window, the target and two counts: the
number of targets the rule explains at that observer and the observer's total number of
targets $s_i$. Its size depends on the number of buckets and not on the trace length.

\emph{Round 2 (merge).} The gateway forms the union $\rules^{\cup}=\bigcup_i \rules_i$
of all proposed rules. For each rule it can compute a weighted average of the coverages
it has received, with each observer weighted by its number of targets,
\[
\hat c(\varphi) \;=\; \frac{\sum_i s_i\,c_i(\varphi)}{\sum_i s_i},
\]
where the sums run over the observers that proposed $\varphi$. This value is only an
estimate. An observer whose local coverage was below $\tau_\ell$, or whose miner failed,
did not propose the rule and sent no counts for it. The gateway therefore applies no
global threshold in this round.

\emph{Round 3 (confirm).} The gateway sends every rule in $\rules^{\cup}$ back to
\emph{every} observer. Each observer checks the rule against its own trace, without
mining, and returns its exact pair of counts. Now the gateway has counts from all
observers, and the coverage of the union, $c^{\cup}$, is exact. The gateway keeps the
rules with $c^{\cup}\ge\tau_g$. This round is what makes Thm.~\ref{thm:recovery} usable
in practice. It costs one more exchange of descriptions and never transfers events.

\paragraph{Topology and message count.}
All messages go between an observer and the gateway. Observers never message each other,
so the communication graph is a star. In Round~1 observer $i$ sends $|\rules_i|$
descriptions. In Round~3 the gateway sends each of the $|\mathcal{C}|$ confirmed
candidates to each of the $N$ observers, and each observer answers with one pair of
counts per candidate. The protocol therefore uses $\sum_i |\rules_i| + 2N|\mathcal{C}|$
messages in total. This is linear in $N$ for a fixed candidate set. It becomes quadratic
in $N$ only when the observers propose disjoint rules, so that $|\mathcal{C}|$ itself
grows with $N$. In our dataset the union has four rules.

\begin{algorithm}[!t]
\caption{\textsc{Mine--Merge--Confirm--Gate}. Rounds 1--3 are
Sect.~\ref{sec:distmine}. Round 4, the gate, is Sect.~\ref{sec:gate}, and the guarantee
and placement choices are Sect.~\ref{sec:degrade}. No raw trace crosses the network.
Observers exchange rule descriptions, count pairs and scalar summaries.}
\label{alg:mmcg}
\footnotesize
\begin{algorithmic}[1]
\Require local traces $\{\trace_i\}_{i\in\obs}$; thresholds
  $\tau_\ell\le\tau_g$; compilation tolerance $\delta_W$; reordering tolerance
  $\delta_R$; loss budget $p^\ast$;
  required detection $\eta$; displacement grid $R=\{0,r_1,\dots,r_{\max}\}$ that the
  transport admits, with $r_{\max}$ its worst case
\Ensure for each accepted rule, a compiled form, a guarantee set and a tier
\Statex \textbf{Precondition.} The $\trace_i$ are separated in time by more than the
  span of every candidate, which is at most $(n{+}1)k$ for an $n$-bucket rule
  (Thm.~\ref{thm:recovery}). If the observers share a clock origin, offset the clocks
  first. Otherwise the merge inflates coverage.
\Statex
\Statex \textit{Round 1: mine, at every observer in parallel}
\For{each $i \in \obs$ \textbf{in parallel}}
  \State $\rules_i \gets \textsc{Miner}(\trace_i, \tau_\ell)$
  \State send $\{\langle \mathrm{desc}(\varphi), c_i(\varphi), s_i(\varphi)\rangle
          : \varphi \in \rules_i\}$ to the gateway
          \Comment{size independent of $|\trace_i|$}
\EndFor
\Statex
\Statex \textit{Round 2: merge, at the gateway}
\State $\rules^{\cup} \gets \bigcup_{i} \rules_i$
\For{each $\varphi \in \rules^{\cup}$}
  \State $\hat{c}(\varphi) \gets
     \bigl(\sum_i s_i(\varphi)\,c_i(\varphi)\bigr) /
     \bigl(\sum_i s_i(\varphi)\bigr)$
     \Comment{estimate only; no threshold}
\EndFor
\State $\mathcal{C} \gets \rules^{\cup}$
     \Comment{confirm the whole union}
\Statex
\Statex \textit{Round 3: confirm, with exact counts from \emph{every} observer}
\For{each $\varphi \in \mathcal{C}$}
  \State send $\mathrm{desc}(\varphi)$ to all; each $i\in\obs$ returns
     $\langle n^{\mathrm{expl}}_i, s_i\rangle$
     \Comment{checking only, no mining}
  \State $c^{\cup}(\varphi) \gets
     \bigl(\sum_i n^{\mathrm{expl}}_i\bigr)/\bigl(\sum_i s_i\bigr)$
     \Comment{exact}
  \State drop $\varphi$ if $c^{\cup}(\varphi) < \tau_g$
\EndFor
\Statex
\Statex \textit{Round 4: gate, plan, place. Each observer computes the same-event
  score, the shuffled scores and the displaced scores for $r\in R$ on its own trace and
  returns scalars. The gateway averages them, weighted by target counts.}
\For{each $\varphi \in \mathcal{C}$}
  \State $W(\varphi) \gets \mathrm{score}(\varphi)-\mathrm{score}(\varphi^{(0)})$
      \Comment{window gain, Eq.~\eqref{eq:wgain}}
  \State $\Delta(\varphi) \gets \mathrm{score}(\varphi) -
      \mathbb{E}_\pi[\mathrm{score}(\varphi,\pi(\trace))]$
      \Comment{shuffle gap, Eq.~\eqref{eq:delta}}
  \If{$W(\varphi) \le \delta_W$}
    \State $\mathit{form} \gets \textsc{stateless}$;\;
           $G \gets \emptyset$ \Comment{Thm.~\ref{thm:gate}}
  \Else
    \State $\mathit{form} \gets \textsc{windowed}$
    \State $D \gets \max_{r\in R}\bigl|\mathrm{score}_r-\mathrm{score}_0\bigr|$
           \Comment{absolute value: disorder can raise or lower}
    \State $G \gets \{\textsc{ordering}\}$ \textbf{if} $D > \delta_R$
           \Comment{not from $|\Delta|$, Prop.~\ref{prop:refute}}
  \EndIf
  \State $(a, m_1,\dots,m_q, \mathit{exact}) \gets
      \textsc{WitnessStructure}(\varphi, \trace)$
      \Comment{largest $a$, smallest $m_\ell$}
  \If{$\mathit{exact}$} \Comment{hypotheses of Thm.~\ref{thm:loss} hold}
    \State $G \gets G \cup \{\textsc{delivery}\}$ \textbf{if}
        $(1-p^\ast)^{a}\prod_{\ell}\bigl(1-(p^\ast)^{m_\ell}\bigr) < \eta$
  \Else
    \State $G \gets G \cup \{\textsc{delivery}\}$; mark $\eta$ \textsc{uncertified}
  \EndIf
  \State place $\varphi$ at the shallowest tier whose evidence path meets
     $p^\ast$ under $G$
\EndFor
\State \Return $\{\langle \varphi, \mathit{form}, G, \mathit{tier}\rangle\}$
\end{algorithmic}
\end{algorithm}

Algorithm~\ref{alg:mmcg} also shows how the three results connect. The merge needs the
traces to be separated in time. The gate takes the merged score and returns a compiled
form and an exposure to reordering. The witness structure turns that into a guarantee
set and a tier. The rest of this section analyses Rounds 1--3.

\subsection{Exact Recovery}
Write $c_i(\varphi)$ for the coverage of rule $\varphi$ at observer $i$, $s_i$ for the
number of target events at $i$, and $c^{\cup}(\varphi)$ for the coverage that a central
miner reports on the union of all traces.

Coverage is a ratio of two counts. If no window can reach from one observer's trace into
another's, both counts of the union are sums of the observers' counts. The coverage of
the union is then a weighted average of the local coverages.

\begin{theorem}[Recovery]\label{thm:recovery}
Suppose every condition of $\varphi$ is evaluated within a single observer's trace, and
the traces are separated in time by more than the span of the rule, that is by more
than $(n{+}1)k$ for an $n$-bucket rule. Then
\begin{equation}\label{eq:weighted}
c^{\cup}(\varphi) \;=\; \frac{\sum_i s_i\,c_i(\varphi)}{\sum_i s_i}
\;\le\; \max_i c_i(\varphi).
\end{equation}
Hence, if $\tau_\ell \le \tau_g$, every such rule that a central miner accepts is proposed
by at least one observer, and \textsc{Mine--Merge--Confirm} recovers it. Two conditions
are needed: at least one proposer's description reaches the gateway, and every trace is
available for checking in Round~3.
\end{theorem}

\begin{proof}
We split the union into the observers' traces. With the separation assumed, every window
lies inside one trace. The number of explained targets of the union is then the sum of
the observers' numbers, and the same holds for the number of targets. This gives the
equality. Note that disjoint traces are not enough by themselves. Traces that are
adjacent in time would let a window reach across the boundary. A separation larger than
$(n{+}1)k$ prevents that. Our offsets separate observers by an hour, against $k\le120$\,s.
A weighted average cannot exceed its largest term. Since $c^{\cup}\ge\tau_g\ge\tau_\ell$,
some observer has $c_i \ge \tau_\ell$ and proposes $\varphi$.

After Round~1 nothing removes $\varphi$ before confirmation. Round~2 forwards the whole
union, and $\varphi\in\rules_j\subseteq\rules^{\cup}=\mathcal{C}$. Round~3 applies
$\tau_g$ to exact counts from every observer. A screen on $\hat c\ge\tau_g$ in Round~2
would fail when a miner fails. An observer whose miner failed contributes nothing to
$\hat c$, however high its coverage. For example, let $\tau_\ell=0.5$ and $\tau_g=0.75$,
and let two observers have equal target counts. One proposes the rule with $c_1=0.60$.
The other has $c_2=1.00$ but its miner failed. Then $c^{\cup}=0.80\ge\tau_g$, while
$\hat c=0.60<\tau_g$. A Round~2 screen would drop a rule that has a valid proposer.
\end{proof}

Thm.~\ref{thm:recovery} has the same structure as the lemma for distributed association
rules by Cheung et al.~\cite{fdm}: any itemset that is frequent globally is frequent at
some site. Coverage plays the role of support, and a time window plays the role of the
itemset. The hypothesis does not carry over. Support adds up over any split of the
data. Coverage adds up only when no window spans two observers. The measurements below
show the practical effect of this difference.

\subsection{Measurement}
\paragraph{What the miner produces.}
We ran \miner\ once per observer, over a shared vocabulary, with \textsc{frag} as the
target. The sets $\rules_i$ are therefore measured, not assumed. They are small. Each
observer produced $0$ to $3$ assertions, $1.38$ on average, and the union has four
distinct rules. For the traffic figures we use a stipulated set of seven candidate rules
at every observer, which gives the conservative $1658\times$ of Prop.~\ref{prop:comm}.
With the miner's own output, Round~3 costs more than Round~1, because confirming a
four-rule union at $21$ observers costs more than proposing it.

Two observations matter more. First, rule \texttt{R1} was proposed independently by $18$
of the $21$ observers, with support between $80.4$ and $90.3$ and correlation between
$89.5$ and $98.9$. The uniform case of Thm.~\ref{thm:recovery} is therefore realised by
the miner itself. Second, the miner \emph{crashed} on $7$ traces, three of them before
writing any output. Local mining is therefore incomplete for a reason unrelated to the
thresholds. This is why Round~2 forwards the whole union. Confirmation in Round~3 adds
the counts of an observer whose miner failed, and \texttt{R1} survives $7$ such failures
after $18$ proposals. These are failures to generate candidates, not failures to read a
trace. Round~3 still needs every trace to be readable by the checker. No round can recover
a rule that no successful miner proposed. One surviving proposer per accepted rule is
therefore required. That is weaker than complete local mining, but it is a real
requirement.

\paragraph{Coverage over all observers.}
Table~\ref{tab:rq1} evaluates five candidate rules over all $21$ observers. The window
gains of \texttt{R1} and \texttt{R2} appear in Table~\ref{tab:rq2}. For rules whose
signal lies inside one packet, the weighted average and the merged value agree closely.
For rules whose antecedent and consequent are different extension headers they differ
sharply. \texttt{R4} has coverage between $2.55\%$ and $22.00\%$ at every single observer,
and its weighted average is $11.64\%$. Naive merging reports $91.59\%$, an inflation of
$79.9$ percentage points. \texttt{R5} behaves the same way, with $78.9$ points.

\begin{table}[t]
\centering\small
\caption{Coverage (\%) over all $21$ observers. \emph{Weighted} is
Thm.~\ref{thm:recovery}. \emph{Naive} merges the traces as recorded. \emph{Offset} merges
them after separating the observers in time. All rules use window $[0{:}4]$\,s.
\textsc{med} abbreviates \textsc{plen\_med}.}
\label{tab:rq1}
\begin{tabular*}{\columnwidth}{@{\extracolsep{\fill}}lrrrr@{}}
\toprule
\textbf{Rule} & \textbf{local range} & \textbf{weighted} & \textbf{naive} & \textbf{offset}\\
\midrule
R1 $\neg$\textsc{med}$\to\neg$\textsc{frag} & 89.7--98.5  & 95.04  & 99.70  & \textbf{95.04}\\
R2 \textsc{med}$\to$\textsc{frag}           & 100.0--100.0 & 100.00 & 100.00 & --\\
R4 \textsc{dst}$\to$\textsc{rt}             & 2.6--22.0   & 11.64  & 91.59  & \textbf{11.64}\\
R5 \textsc{rt}$\to$\textsc{frag}            & 4.0--25.2   & 16.58  & 95.50  & \textbf{16.58}\\
R6 \textsc{hlim}$\to$\textsc{frag}          & 96.7--100.0 & 99.35  & 100.00 & --\\
\bottomrule
\end{tabular*}
\end{table}

The cause is the shared clock origin of Sect.~\ref{sec:setting}. Merging puts $21$
traces on one time axis. A window opened by an antecedent at one receiver can then be
closed by a consequent recorded on another continent. The fix is to meet the hypothesis
of Thm.~\ref{thm:recovery}. We offset each observer by the span of the previous one plus
one hour. The merged value then equals the weighted average, with residuals of $-0.000$,
$+0.000$ and $+0.000$ points on \texttt{R1}, \texttt{R4} and \texttt{R5}.
Thm.~\ref{thm:recovery} is therefore not an approximation of central mining. It states
central mining correctly. Naive merging is an error that is easy to miss, because it
raises the reported score.

\begin{proposition}[Communication]\label{prop:comm}
Including the confirmation round, the protocol exchanges
$O\!\left(\bigl(\sum_i |\rules_i| + N|\mathcal{C}|\bigr)\, n_{\max}\right)$
symbols, where $n_{\max}$ is the largest number of buckets in a candidate rule. This is
independent of trace length. Merging the traces costs $O(\sum_i |\trace_i|)$.
\end{proposition}
\begin{proof}
In Round~1, each locally mined rule costs one description of $O(n_{\max})$ symbols plus
two counts, and there are $\sum_i|\rules_i|$ such rules. Round~2 happens at the gateway.
In Round~3, each of the $|\mathcal{C}|$ candidates is sent to all $N$ observers, and each
returns a pair of counts. This costs $O(N|\mathcal{C}|n_{\max})$ symbols. No round sends
events. Hence no term depends on $|\trace_i|$. Merging sends every event.
\end{proof}

The term $N|\mathcal{C}|$ is the price of exactness. The saving holds whenever the
candidate set is small compared with the traces, which is the usual situation in mining.
With seven stipulated candidates per observer and $N=21$, Round~1 sends $10{,}101$\,B of
descriptions. Round~3 sends the same amount and receives $1{,}176$\,B of count pairs. The
total is $21{,}378$\,B, against $35{,}435{,}638$\,B for the traces. Round~1 alone is
$3508\times$ smaller. We quote $1658\times$ for the full protocol, because only the full
protocol computes an exact answer.
\section{The Compilation Gate}\label{sec:gate}

A miner writes its rules with temporal operators. That does not show that the window
does any work. Two separate questions arise. The first is whether a check on single
events, with no window and no state, would give the same score. If so, the window and
its state are wasted. The second is how much the score depends on the order of events
inside the window, because order is what the transport disturbs. Each question gets its
own statistic. We also show a real rule for which using the wrong statistic would have
been costly.

\subsection{Two Statistics}
\paragraph{Shuffle gap.}
Let $\pi$ be a \emph{joint-row shuffle}. It permutes the per-event attribute tuples as
whole units over the fixed list of timestamps. Every association inside an event and
every marginal distribution is kept. Only the order of events changes. The \emph{shuffle
gap} is
\begin{equation}\label{eq:delta}
\Delta(\varphi) \;=\; \mathrm{score}(\varphi,\trace) \;-\;
   \mathbb{E}_\pi\!\left[\mathrm{score}(\varphi,\pi(\trace))\right].
\end{equation}
A negative $\Delta$ means that shuffling raises the score, and a positive $\Delta$ means
that it lowers it. $|\Delta|$ is the size of the effect in either case.

A gate that compiles a rule to a stateless check whenever $\Delta$ is small would drop
rules that need their window. Consider a rule whose antecedent and target never hold in
the same event, with a window wide enough that almost every target has some antecedent
inside it. Shuffling barely changes its score, and $\Delta\approx0$. Its stateless version
requires antecedent and target in one event. That never happens, and its score is zero. A
small $\Delta$ says that order carries little information. It does not say that the
window carries none. Saturated rules are exactly this case (Sect.~\ref{sec:fp}).

\paragraph{Window gain.}
The gate must therefore compare a rule with its stateless version. The \emph{same-event
version} $\varphi^{(0)}$ of $\varphi$ sets every window to $[0{:}0]$. All buckets must then
hold in one event. The \emph{window gain} is
\begin{equation}\label{eq:wgain}
W(\varphi) \;=\; \mathrm{score}(\varphi) - \mathrm{score}(\varphi^{(0)}).
\end{equation}

\begin{theorem}[Gate soundness]\label{thm:gate}
Let the score be coverage. If $W(\varphi)\le\delta_W$, then the per-event check
$\varphi^{(0)}$ has a coverage at most $\delta_W$ below that of $\varphi$. In other
words, window state adds at most $\delta_W$ to the score.
\end{theorem}

\begin{proof}
For positive existential rules, $\varphi^{(0)}$ is a stronger condition than $\varphi$.
Any witness of $\varphi^{(0)}$ is a witness of $\varphi$ in which all buckets fall on one
event. Hence $\mathrm{score}(\varphi^{(0)}) \le \mathrm{score}(\varphi)$ and
$W(\varphi)\ge0$. The rule $\varphi^{(0)}$ needs all buckets at one event. It can therefore be
evaluated without window state. By assumption its score is at most $\delta_W$ lower than
that of $\varphi$. The argument needs coverage. With confidence as the score, a stronger
antecedent can move the conditional in either direction.
\end{proof}

The guarantee holds for the trace and score on which $W$ was computed. We therefore test
the decision on held-out data. We fixed two splits before computing $W$. The first
divides $12$ observers into two groups of six. The second divides time into a first and
a second half. Identifiers, seeds and per-rule values are in the artifact. In both
splits, every decision taken over the pooled observers held on the other part. The one
rule compiled stateless, \texttt{R2}, has held-out $W=0.00$. No windowed rule has a
held-out $W$ below $\delta_W$. Decisions taken at single observers did not transfer. At
one observer $W$ moved by up to $28.3$ points between halves, and two rules would have been
compiled stateless although their held-out $W$ exceeds $\delta_W$. This supports taking
the decision at the gateway.

The gate keeps the two statistics apart. $W$ decides \emph{compilation}, that is,
whether the window is worth its state. $\Delta$ measures sensitivity to the shuffle,
which keeps the timestamps in place and moves only attributes.
Section~\ref{sec:degrade} shows that $\Delta$ does not bound the effect of moderate
displacement. A rule with $W\le\delta_W$ compiles to a per-event check, with no window
that could be disordered. A rule with $W>\delta_W$ keeps its window. Its ordering
requirement is then decided by
$D=\max_{r\in R}|\mathrm{score}_r-\mathrm{score}_0|$ over the range of displacement that
the deployment admits. A large $|\Delta|$ may not be reached inside that range, and a
small $|\Delta|$ is no evidence of safety (Prop.~\ref{prop:refute}).

\subsection{Measurement}
\begin{table}[t]
\centering\small
\caption{Window gain $W$ and shuffle gap $\Delta$ (points of coverage). $W$ decides
compilation. $\Delta$ reports sensitivity to the shuffle only. \texttt{R6} is the key case.
Its $\Delta$ interval contains zero while its $W$ interval is near $78$. A single
statistic would misclassify it.}
\label{tab:rq2}
\setlength{\tabcolsep}{3pt}
\begin{tabular}{@{}lccll@{}}
\toprule
\textbf{Rule} & \textbf{$W$ [95\% CI]} & \textbf{$\Delta$ [95\% CI]} &
\textbf{compile} & \textbf{guar.}\\
\midrule
R1 & 4.20 [2.39, 6.01]    & $-3.58$ [$-5.5$, $-1.7$]  & windowed & warn\\
R2 & \textbf{0.00} [0.00, 0.00] & 0.00 [0.00, 0.00]    & \textbf{stateless} & --\\
R4 & 15.63 [9.10, 22.2]   & $-83.9$ [$-90.5$, $-77.4$] & windowed & warn\\
R5 & 18.17 [10.8, 25.6]   & $-81.3$ [$-88.7$, $-74.0$] & windowed & ord.\ /$R$\\
R6 & \textbf{78.37} [76.1, 80.7] & $\mathbf{-0.31}$ [$-1.23$, 0.61] & windowed
   & none /$R$\\
\midrule
\multicolumn{5}{@{}p{\columnwidth}@{}}{\footnotesize Six observers, fifteen shuffles each
over the \emph{full} attribute tuple of \eqref{eq:delta}; $\delta_W=\delta_R=1$.
Student-$t$ intervals with $5$ degrees of freedom across observers. The within-observer
shuffle standard deviation is at most $0.101$. \emph{warn} marks a large gap for which no
displacement sweep was run. We make no deployment claim for these. The guarantee comes from $D$
(Sect.~\ref{sec:degrade}). Only \texttt{R5} and \texttt{R6} were swept.}\\
\bottomrule
\end{tabular}
\end{table}

Table~\ref{tab:rq2} reports both statistics. The main rule of the dataset, \texttt{R2},
has $W = 0.00$. Its window adds nothing over its same-event version, and by
Thm.~\ref{thm:gate} it compiles to a stateless check. This turns the earlier negative
finding on this dataset~\cite{csr} into a decision procedure.

\texttt{R6} shows why the two statistics must be kept apart. Its window gain is
$W=78.37$ $[76.1,80.7]$, while its shuffle gap $\Delta=-0.31$ $[-1.23,0.61]$
\emph{contains zero}. The window is essential, yet the ordered and shuffled scores cannot
be told apart. We do not conclude that order is irrelevant for \texttt{R6}. The rule is
almost saturated at $99.24$, which limits how far any reordering can move it. The
evidence is therefore weaker than the interval suggests. A gate based on $\Delta$, as in our earlier
formulation, would have compiled \texttt{R6} to a stateless check and lost almost eighty
points of coverage. As an instance of Algorithm~\ref{alg:mmcg}, \texttt{R5} has
$W=18.2>\delta_W$ and is compiled as windowed. Its $D$ over the swept range is $33.3$
points, which puts \textsc{ordering} in its guarantee set. \texttt{R2}, \texttt{R6} and
\texttt{R1s} (a variant of \texttt{R1} with a zero-width window) have $D\le0.5$ and request no ordering guarantee.

Shuffling almost always \emph{raises} the score. In a randomly ordered dense trace, some
event easily falls inside a window. Section~\ref{sec:degrade} starts from this
observation.
\section{Loss, Reordering, and Induced Distortion}\label{sec:degrade}

A rule that passes mining and compilation can only be as good as the evidence that
reaches it. The two transport effects that cannot be avoided behave very differently.
Loss has a closed form that depends on the traffic, not on the rule's syntax. Reordering
has no such form. It lowers scores as easily as it raises them, its worst case lies
between small and large displacement, and the cheap shuffle statistic measures something
else.

\subsection{Detection Under Evidence Loss}
A simple estimate says that a rule which needs $k+1$ events to witness a violation
detects it with probability $(1-p)^{k+1}$ under independent per-event loss $p$. For
existential windows this is wrong, and the error is measurable. Our windowed rules lose
detections like $(1-p)^1$, not like $(1-p)^2$. A window usually holds many interchangeable
candidates, and all of them are lost together only with probability $p^m$.

\begin{definition}[Witness structure]
Fix a rule and a violation. Its witness consists of $a$ \emph{irreplaceable} events,
which the rule pins down uniquely, and, for each existential window $\ell$, a set
$M_\ell$ of $m_\ell$ \emph{candidates}, any one of which completes that window. The target
event is always irreplaceable. If it is lost, the detection is lost whatever happens to
the antecedents. An antecedent bucket is irreplaceable only when its window happens to
hold exactly one candidate.

The witness structure belongs to a violation. A rule therefore has one per target. When we quote
a single $(a,m_1,\dots,m_q)$ we take the largest $a$ and the smallest $m_\ell$ over the
targets. This makes \eqref{eq:loss} a conservative guarantee. When we compare with
measured averages we use the dataset mean
$|T|^{-1}\sum_{t}(1-p)^{a_t}\prod_\ell(1-p^{m_{t,\ell}})$.
\end{definition}

\begin{theorem}[Detection under loss]\label{thm:loss}
Let a violation of $\varphi$ have $a$ irreplaceable witness events and $q$ existential
windows with candidate sets of sizes $m_1,\dots,m_q$. Assume that (i) the $a$ events and
the $q$ candidate sets are pairwise disjoint, (ii) they are \emph{freely combinable}, that
is, any choice of one member from each candidate set, together with the $a$ events, is a
witness, and (iii) they are \emph{exhaustive}, that is, the violation has no other
witness. Under independent per-event loss with probability $p$,
\begin{equation}\label{eq:loss}
\mathrm{detect}(p) \;=\; (1-p)^{a}\prod_{\ell=1}^{q}\bigl(1 - p^{m_\ell}\bigr).
\end{equation}
\end{theorem}

\begin{proof}
Detection needs every irreplaceable event to arrive. By independence this has probability
$(1-p)^a$. Each window must also keep at least one candidate. This fails only if all
$m_\ell$ candidates are lost, which has probability $p^{m_\ell}$. Disjointness makes the
survival of different sets independent. Free combinability makes detection the
\emph{intersection} of these events. The $q+1$ factors therefore multiply. The rule is
positive in its buckets. Loss can therefore remove a detection but never create one.
\end{proof}

Free combinability is as important as disjointness. Without it the formula fails in the
\emph{unsafe} direction, because consecutive links of a chain can be coupled. A survivor
in one window is useful only together with a compatible survivor in the next. Take $k=1$,
a target at $0$, first-window candidates at $-0.1$ and $-0.9$, and second-window
candidates at $-0.15$ and $-1.5$. Each second-window candidate is compatible with exactly
one first-window candidate. Detection is then $(1-p)(2(1-p)^2-(1-p)^4)$, which is lower
than \eqref{eq:loss}. We measure this case below. Overlap between windows pushes the other
way, because shared events create positive dependence. So \eqref{eq:loss} is exact under
its assumptions, and without them it is not a bound in either direction. This is
important because Algorithm~\ref{alg:mmcg} uses it to decide whether to request a delivery
guarantee.

Disjointness is what turns \eqref{eq:loss} into a product. When windows share events, the
survival events of the windows become \emph{positively} dependent. The product of the
marginal probabilities then \emph{under}-states detection. Two windows that share one
candidate survive together with probability $1-p$, while the product gives
$(1-p)^2\le 1-p$. For monotone survival events under independent loss the product is a
lower bound in this case. An exact value would need the reliability polynomial of the
witness hypergraph. Our chain rules satisfy disjointness because of the projection we use,
not because of IPv6. The schema records one \texttt{nxt} value per packet. The
\textsc{has\_*\_eh} conditions are therefore mutually exclusive within one event, and no
event can fill two buckets. A projection that kept the full header chain would lose this
property. The chain results would then be lower bounds. We do not treat the general
overlapping case.

Equation~\eqref{eq:loss} contains the length-based formula as the special case $q=1$,
$m_1=1$, which gives $(1-p)^{a+1}$. It also explains why real windowed rules behave like
$(1-p)$. In a dense trace no antecedent window holds exactly one candidate. The target is
then the only irreplaceable event, and $a=1$.

We validated all three parameters on planted witnesses. Sweeps over $m$ at $a=1$, and over
$a$ against $m$, agree with the law to within $0.0155$ and $0.0098$. Both sweeps have
$q=1$, and the dataset cannot test the product either, because window occupancies of $36$
to $565$ put every factor within $10^{-10}$ of one. We therefore planted explicit
\emph{timestamped} chains with small $m$, dropped events independently, and decided
detection by a backward search for a witness among the surviving events. The geometry of
the trace, not \eqref{eq:loss}, fixes the outcome in this test. Over
$q\in\{1,2,3\}$ and $m\in\{1,2,3\}$, with $200{,}000$ blocks per cell, the law holds to
within $0.0025$, against a binomial standard error of $0.0011$. The test discriminates. At
$p=\tfrac12$ these cells range from $0.0625$ to $0.4375$. The coupled chain measures
$0.2190$. This matches its own closed form ($0.2188$) and not \eqref{eq:loss} ($0.2812$).
The overstatement grows from $1.7\%$ at $p=0.1$ to $28.6\%$. Free combinability is the
dividing line, and we measured both sides of it.

\paragraph{Occupancy decides fragility, length does not.}
The template \eqref{eq:template} allows any number of antecedent buckets. Longer chains
might be expected to be more fragile. They are not. We implemented the full $n$-bucket
template, which the published validator cannot express. Chains of one, two and three
buckets over sparse cross-family flags at $k=120$\,s degrade \emph{identically}. They track
$(1-p)$ to within $0.005$ at $p\in\{0.10,0.20,0.30,0.50\}$. Each link is existential, and
the measured occupancy of $36$ to $565$ candidates puts $(1-p^{m})\approx1$ at every link,
with $a=1$ throughout. Occupancy is not a property of the trace alone. At one fixed
observer, changing only the rule moves the median occupancy from $0$ to $58$ at $k=4$,
and $a$ takes both values $1$ and $2$. The pair of rule and trace fixes $a$, and the gate
has to measure it. Thinning confirms the mechanism. We keep each non-target event with
probability $\theta$. This scales occupancy and leaves the rule and its targets unchanged.
At $p=0.5$, detection stays at the linear $(1-p)$ value ($0.501$, $0.508$, $0.495$,
$0.492$) while the median occupancy falls from $372$ to $109$, $42$ and $12$. It then moves
toward the quadratic value at $m=4$ ($0.444$ against the law's $0.469$) and at $m=1$
($0.352$ against $0.250$). The remaining gap comes from evaluating the law at the
\emph{median} of a skewed distribution.

Natural traffic lies beyond the end of this axis. The MAWI backbone capture of
31 December 2025 carries $508$ extension-header packets among $8{,}246{,}274$ IPv6 frames,
a prevalence of $0.0062\%$. It has no Routing, Destination Options, ESP or AH headers at
all. This is consistent with the reported dropping of packets with extension
headers~\cite{rfc7872}. Windows in that trace are empty, not just sparse. It yields no
cross-family witnesses and cannot measure detection. This is why the behaviour is
probed actively. It also limits where such rules can be deployed at all.

On JAMES, relative detection at $p\in\{0.1,0.3,0.5\}$ is $0.899$, $0.701$ and $0.501$
for the same-event rule, and within $0.005$ of these for both windowed rules. All of
them track $(1-p)$, as \eqref{eq:loss} predicts at $a=1$ and $m\gg1$. A length-based
estimate would give $0.810$, $0.490$ and $0.250$ and cannot explain the data.

\subsection{Distortion Under Reordering}
Reordering can remove detections or invent them. Which effect dominates depends on where
a rule's evidence sits relative to its window, not on the transport. For the network and
industrial rules we measured, the dominant effect is inflation. Moving timestamps brings
extra candidates into windows and raises the score.

Additive displacement and the shuffle of Sect.~\ref{sec:gate} are different
perturbations. Displacement adds a random offset to each timestamp and re-sorts the
events. Large displacement eventually spreads events so far apart that no two lie within
one window. The shuffle keeps the timestamps in place and therefore keeps the density of
events in time. Displacement taken to its limit might be expected to give the
shuffled score. It does not, because our rules are bounded in \emph{time}, not in position.

\begin{proposition}[Displacement end points]\label{thm:inflate}
Let the displacements be $\varepsilon_i = r Z_i$, where the $Z_i$ are independent,
continuous and have a bounded density. Let $\mathrm{score}_r$ be the expected score after
displacement and re-sorting, on a finite trace with fixed window $k$. Our experiments use
$Z_i \sim U[-1,1]$, so that $\varepsilon_i \sim U[-r,r]$. Then
$\mathrm{score}_0=\mathrm{score}(\varphi,\trace)$ and
$\lim_{r\to\infty}\mathrm{score}_r=\mathrm{score}(\varphi^{(0)})$, the score of the
same-event version. The total change from one end point to the other is exactly
$-W(\varphi)$.
\end{proposition}

\begin{proof}[Proof sketch]
Under large displacement, coincidences disappear. For two distinct events the displaced
gap has density $O(1/r)$. A union bound then shows that, with probability tending to one,
all events are more than $k$ apart. Only witnesses inside single events remain, and these
are the witnesses of $\varphi^{(0)}$. Appendix~\ref{app:proofs} gives the details.
\end{proof}

This limits how $\Delta$ may be used. The shuffle holds the timestamp grid \emph{fixed}.
It keeps the density that makes matches across events possible. Displacement destroys
that density. For a logic with bounded durations the two are therefore not equivalent
in the limit. $\Delta$ is best read as a cheap test of whether events are exchangeable.
On the synthetic trace \emph{tight}, described below, the shuffle gives $0.248$ and
displacement at $r=10^5$ gives $0.002$.

Prop.~\ref{thm:inflate} fixes the two end points of the displacement curve. It says
nothing about the curve between them. A natural conjecture is that
$\mathrm{score}_r-\mathrm{score}_0\le|\Delta(\varphi)|$ for every $r$, so that the
shuffle gap bounds the effect of any displacement. This would follow if $\mathrm{score}_r$
were monotone in $r$. A bounded displacement could, however, move candidates into the
relevant windows more effectively than a fully random one, and exceed $|\Delta|$. The
conjecture is in fact false.

\begin{proposition}[The shuffle gap does not bound displacement]\label{prop:refute}
There are traces and positive existential rules for which $\mathrm{score}_r$ is not
monotone in $r$ and for which $|\mathrm{score}_r-\mathrm{score}_0| > |\Delta(\varphi)|$.
The trace \emph{offset} below violates the one-sided form
$\mathrm{score}_r-\mathrm{score}_0\le|\Delta|$. The trace \emph{tight} shows a decrease
larger than $|\Delta|$.
\end{proposition}

\begin{proof}[Proof sketch]
Take a trace of well-separated pairs of antecedent and target, each at distance
$\tfrac32k$. Its score is $\mathrm{score}_0=0$. No two of its timestamps lie within $k$,
so the shuffled score is also zero and $\Delta=0$. Displacement with $r>k/4$ nevertheless
covers some pairs with positive probability. Appendix~\ref{app:proofs} gives the details.
\end{proof}

Two synthetic traces with $400$ targets, $k=1$ and $40$ repetitions quantify this. We
sweep $r$ over $0$, $0.5$, $1$, $2$, $5$ and $10^5$. In \emph{tight}, each window holds one
antecedent just inside it. Its score runs $1.000$, $0.750$, $0.433$, $0.229$, $0.098$ and
$0.002$, against a shuffled score of $0.248$. The curve is monotone and ends at the
same-event score, not at the shuffled score. In \emph{offset}, the antecedent sits just
\emph{outside} each window. Its score runs $0.000$, $0.120$, $\mathbf{0.248}$, $0.187$,
$0.098$ and $0.003$, against a shuffled score of $0.000$. It peaks near $r\approx k$ and
then decays. Here $24.8$ points of inflation occur where $|\Delta|=0$ predicts none.

The mechanism is a resonance. Inflation is largest when the displacement scale is close to
the distance between where the evidence sits and where the window expects it. Beyond that
scale, disorder destroys structure faster than it creates matches. Both small and large
displacement can therefore be safer than moderate displacement. A monotone bound would
predict the opposite.

\paragraph{What the shuffle gap is good for.}
Equation~\eqref{eq:delta} is cheap, and the gate computes it anyway. It is useful for
deciding which rules to sweep first. It is not the end point of additive displacement and
not a bound on the distortion a transport can cause. It is weakest against a
\emph{chosen} reordering. On \emph{tight}, a greedy adversary reaches $1.000$ against a
shuffled score of $0.248$. Choosing an ordering guarantee therefore requires the whole
curve. One must sweep $r$ over the range the transport can produce and take the maximum,
including values of $r$ close to $k$. Prop.~\ref{prop:refute} places the peak there, and
a coarse grid would miss it.

The sweep on real data shows the inflating direction clearly. Over six observers with
five repetitions per cell, coverage under displacement by $r\in\{0,1,2,4,8,16\}$\,s at
$k=4$ runs $18.18$, $19.78$, $21.71$, $25.86$, $34.60$ and $51.45$ for \texttt{R5}. For
\texttt{R6} it runs $99.24$, $99.40$, $99.56$, $99.73$, $99.68$ and $99.76$.
\texttt{R2} ($100.00$) and \texttt{R1s} ($92.20$) stay constant to two decimals.
\texttt{R5}'s ordered coverage of $18.18$, against $0.01$ for its same-event version, gives
the $W=18.17$ of Table~\ref{tab:rq2}. \texttt{R5}, with $|\Delta|\approx81$, therefore
inflates to $2.83\times$ its ordered coverage at $r=16$\,s. It is not heading for the
shuffled score. By Prop.~\ref{thm:inflate}, displacement drives the score to the
same-event value, $0.01$ for \texttt{R5}. The curve must therefore turn down beyond the
range that a wide-area path would produce.

We could not test the protective direction cleanly on this dataset. Every rule with a
small $|\Delta|$ is confounded. \texttt{R2} sits at the ceiling of $100.00$, where no
reordering can move it. \texttt{R6} is almost saturated at $99.24$. \texttt{R1s} has a
zero-width window and ignores order by construction. All three are flat. This is
consistent with low exposure but does not establish it, and it does not imply that the
original windowed rules would stay flat on a future trace. The one unconfounded case is
synthetic and goes the other way. \emph{offset} has $\Delta=0$ and inflates by $24.8$
points.

\subsection{From Score Distortion to False Alarms}\label{sec:fp}
Score inflation is a problem only if it leads to wrong decisions. The WUSTL-IIoT-2021 dataset
is labelled per flow. A firing can therefore be classified. If its target is an attack it
is a true positive, and if its target is normal traffic it is a false positive.
Table~\ref{tab:rq8} reports what reordering does over $200{,}000$ flows, of which $7.2\%$
are attacks.

\begin{table}[t]
\centering\small
\caption{Reordering on labelled IIoT flows ($200{,}000$ flows, $k=10$\,s). The rules are
\texttt{F1} $\textsc{loss}{>}0\to\textsc{rate}{>}500$, \texttt{F2}
$\textsc{dur}{<}0.1\to\textsc{bytes}{>}2000$ and \texttt{F3}
$\textsc{rate}{>}500\to\textsc{loss}{>}0$. A prime marks the same rule after thinning
non-target events to $\theta=0.02$, which removes saturation. \emph{sat.} is the fraction
of targets already firing, and $\Delta$ is the shuffle gap of the firing rate. FDR is
$\mathrm{FP}/(\mathrm{TP}{+}\mathrm{FP})$, a false \emph{discovery} rate over firings.
It is not a false-positive rate over negatives; see Axelsson~\cite{baserate}.}
\label{tab:rq8}
\setlength{\tabcolsep}{4pt}
\begin{tabular}{@{}lrrcrrrr@{}}
\toprule
& & & & \multicolumn{4}{c}{FP inflation at $r=$}\\
\cmidrule(l){5-8}
\textbf{Rule} & \textbf{sat.} & \textbf{$\Delta$} & \textbf{FDR $0\!\to\!120$}
 & \textbf{1} & \textbf{5} & \textbf{30} & \textbf{120}\\
\midrule
F1        & 1.00 & $0.0$   & $0.233\to0.233$ & 1.000 & 1.000 & 1.000 & 1.000\\
F1$'$     & 0.98 & $-1.3$  & $0.218\to0.218$ & 0.985 & 0.985 & 0.992 & 0.996\\
F2        & 1.00 & $0.0$   & $0.195\to0.195$ & 1.000 & 1.000 & 1.000 & 1.000\\
F2$'$     & 0.69 & $-10.6$ & $0.228\to0.201$ & 0.962 & 0.962 & 0.962 & 0.970\\
F3        & 0.49 & $-49.5$ & $0.879\to0.886$ & \emph{0.943} & \emph{0.971} & 1.046 & 1.072\\
F3$'$     & 0.19 & $\mathbf{-78.6}$ & $\mathbf{0.687\to0.746}$
 & \emph{0.946} & \emph{0.969} & 1.083 & \textbf{1.339}\\
\midrule
\multicolumn{8}{@{}p{\columnwidth}@{}}{\footnotesize Italic cells are \emph{below} $1$:
false positives first fall and then rise. Non-integer counts are means over
repetitions.}\\
\bottomrule
\end{tabular}
\end{table}

At the largest displacement, inflation happens to grow with $|\Delta|$ across these three
rules: $1.00\times$ at $1.3$, $1.07\times$ at $49.5$ and $1.34\times$ at $78.6$. This is an
association on three points and not a bound. Prop.~\ref{prop:refute} shows that no such
bound exists. Even this association depends on reading only the end points. \emph{Every}
unsaturated cell is non-monotone in $r$. False positives first \emph{fall}, by $5.4\%$ for
\texttt{F3}$'$ and by $3.8\%$ for \texttt{F2}$'$. The peak between the end points is
therefore not a synthetic curiosity. Labelled industrial traffic shows it too, and this
is why Algorithm~\ref{alg:mmcg} decides from $D$ over a grid. The two saturated rules stay
flat, as expected when there is no room for additional firings.

The inflation consists \emph{entirely} of false alarms. For \texttt{F3}$'$ the number of
true positives changes by less than $0.03\%$, while false positives grow by $8{,}522$. The
FDR worsens from $0.687$ to $0.746$. Disorder finds no additional attacks. It produces
additional alarms. Merging the source hosts onto one time axis reproduces the effect at
$1.13\times$. WUSTL's hosts share a clock. We therefore created the JAMES condition by resetting
each host's clock to zero.

One case goes the other way. Table~\ref{tab:rq8} lists all six cells. The reader can therefore
check that we did not select the direction. \texttt{F2}$'$, with $|\Delta|=10.6$, gains
$286$ \emph{true} positives under reordering while its false positives fall by $18$. Its
FDR \emph{improves} from $0.228$ to $0.201$.

What decides the direction is the mix of labels in the windows that a rule's candidates
fall into, and nothing in our analysis predicts that mix. $|\Delta|$ therefore determines
neither the direction of the change nor its size at intermediate displacement. We claim
only what it is, a cheap test of exchangeability on a fixed grid, and not a predictor of
harm.

\subsection{Measured Broker Behaviour}\label{sec:envelope}
All results so far inject $p$ and $r$ directly into the traces. To check them against a
real transport, we ran the brokers of the fabric. MQTT with QoS~1 carries ingest, and
NATS JetStream with durable streams carries traffic from the gateway to the back end. The
brokers ran in containers on one host. Each measurement used $2000$ messages per hop,
three runs, a $4$\,ms pace and one publisher on one connection. Versions and host
configuration are in the artifact.

\paragraph{Loss is severe, silent and not independent.}
We published a burst before the consumer attached. This models a monitor that lags or
recovers and replays the durable stream. Any message evicted from the stream is then
evidence the monitor never sees. A consumer that keeps pace would see the events first.
When $2000$ events are published into a stream bounded at $500$ messages, $25\%$ are
kept. With a bound of $1000$, $50\%$ are kept. Under the \emph{discard-old} policy not a
single publish returns an error. Continuity is lost without any signal. Under
\emph{discard-new} the same volume is rejected and every drop is reported. The overload
ratio sets how much is lost, and the configuration sets whether the loss is visible.

Eviction is not independent loss. The broker kept a \emph{contiguous suffix}, the messages
with identifiers $1500$ to $1999$. We applied this exact pattern to the real trace and
measured the result. Here only non-target events are dropped, so the target always arrives; in the random-loss measurements above it could be lost too. At the same retained fraction of $0.25$, relative detection under
independent retention is $0.98$ to $0.99$ for \texttt{R1} and \texttt{R5} on two observers
(means over $20$ draws, each draw between $0.965$ and $0.995$). A window with many
candidates survives random thinning. Under the broker's own pattern, relative detection is
$0.23$ to $0.51$. That is a gap of up to $0.76$ in the unsafe direction. The shape of the
loss costs more than its rate. Here \eqref{eq:loss} serves as a benchmark for independent
loss, not as a prediction. The delivery guarantee is the one that counts here.

\paragraph{No reordering was observed.}
Latency was $0.85$\,ms at the median and $2.34$\,ms at the $99$th percentile on
JetStream, and $0.88$ and $1.97$\,ms over MQTT. The widest range over three runs of $2000$
events per hop was $27.9$\,ms. An \emph{unordered} transport with this spread could swap
adjacent publications sent $4$\,ms apart. The tested single-connection path nevertheless
kept the order exactly. On both hops we counted \textbf{zero} inversions and zero
displacement in rank. A spread of delays is therefore not the same as displacement. We
take the observed range as a conservative $\widehat{r}_{\max}=27.9$\,ms. This is $143$
times smaller than the window $k=4$\,s. Over the grid $[0,\widehat{r}_{\max}]$, $D$ is at
most $0.018$ points for all four rules, more than $55$ times below $\delta_R=1$.
Algorithm~\ref{alg:mmcg} therefore requests no ordering guarantee for any \emph{tested}
rule in this configuration. This is an empirical estimate and not a property of the
transport. Several publishers, partitioned subjects or a wide-area path could raise it.
The \emph{offset} construction of Prop.~\ref{prop:refute} peaks near $r\approx k$, more
than $140$ times beyond the observed range. In this deployment, loss is the binding effect and ordering
is not, and neither conclusion follows from $|\Delta|$.

\subsection{Comparing the Perturbations}
Naive merging, timestamp displacement and the shuffle are \emph{different}
perturbations. Each one weakens the link between an event's position in time and its
attributes, and each can therefore remove or invent matches. Four of our measurements
show the same phenomenon: the $79.9$-point merging inflation, the $2.83\times$ inflation
under reordering, the shuffle gap of $-83.9$ points, and the $8{,}522$ invented alarms. The
main failure mode of mining followed by monitoring is a distortion of temporal correlation
caused by collection and transport. In the security cases we measured, that distortion
mostly invents matches.

\paragraph{Implications for deployment.}
A fabric offers a menu of guarantees at different costs~\cite{fabric}. The analysis says
which of them a mined rule needs. Under the assumptions of Thm.~\ref{thm:loss}, the
delivery guarantee can be priced exactly from the witness structure. Without those
assumptions, Algorithm~\ref{alg:mmcg} requests it as a precaution, and $\eta$ is then not
certified. The ordering decision comes from $D$ over the admitted range, measured in
Sect.~\ref{sec:envelope}. Against an adversary who \emph{chooses} the displacement,
neither quantity is sufficient.
\section{Threats to Validity}\label{sec:threats}

\paragraph{Loss and displacement models.}
Thm.~\ref{thm:loss} assumes independent loss. Section~\ref{sec:envelope} shows the cost
of that assumption. Real eviction is contiguous and removes up to $0.76$ more relative
detection than the independent law predicts at the same rate. We also model displacement
as independent symmetric jitter followed by re-sorting. Real transports produce one-sided,
correlated delay and keep the order within each partition.

\paragraph{Saturation.}
Saturated rules have little room for \emph{additional} firings, and $\Delta\approx0$ for
them. They are not immune, because reordering can remove existing matches. Two of our
three IIoT rules are saturated at full density. We had to thin the traces to see any
effect at all.

\paragraph{The shuffle control.}
The shuffle gap inherits the limits of its control. A joint-row shuffle keeps the
structure inside each event. A rule could therefore depend on order in a way the shuffle does
not reveal. The \emph{compilation} criterion does not use the shuffle at all. An earlier
version of the control left the \texttt{hlim} and \texttt{flow} attributes in place, so
the antecedent of \texttt{R6} was not perturbed. Over the full tuple, $\Delta(\texttt{R6})$
moves only from $-0.35$ to $-0.31$ and its interval still contains zero. The
classification of \texttt{R6} survives the change of control, and saturation remains the
only confound.

\paragraph{Thresholds.}
$\delta_W$ and $\delta_R$ are free parameters. The within-observer standard deviation of
the shuffle is at most $0.101$ points. The widest interval of $\Delta$ across observers
spans about $15$ points. Setting $\delta_W=\delta_R=1$ is therefore conservative against
shuffle noise by an order of magnitude. It is \emph{not} conservative against the spread
between observers. With more heterogeneous observers the thresholds should be set from
that spread.

\paragraph{Scope of recovery.}
Thm.~\ref{thm:recovery} covers rules evaluated within one observer, whose counts add up
across traces separated in time. Rules with antecedent and target at different observers
are \emph{outside the scope} of the protocol. It exchanges descriptions and counts and
never events. It cannot assemble such a witness. Supporting these rules would need
clocks that agree within a bounded skew and an exchange of timestamped events. We do not
attempt this.

\paragraph{Causality.}
Mined rules report temporal precedence and association. They do not report the effect
of an intervention, and nothing here supports a causal reading~\cite{csr}.

\paragraph{Evidence limitations.}
Section~\ref{sec:envelope} measures brokers that run on one host with one publisher,
where order is kept exactly. Several publishers, partitioned subjects or a wide-area path
would all allow displacement, and we emulate none of them. We therefore treat
$\widehat{r}_{\max}$ as a baseline for this configuration and not as a bound for others.
We do not measure the savings in state, latency or bandwidth that compilation is meant to
deliver. The loss law is validated on two of the $21$ observers, and the occupancy axis
on a single pair of antecedent and target. The planted sweeps carry the coverage of the
parameters. Finally, our dataset comes from an active-probing campaign whose vocabulary
almost never appears in backbone traffic. Its rules are instruments for studying the
mechanism, not rules an operator would deploy. What transfers is the structure of
existential windows with bounded duration over per-event conditions.

\section{Related Work}\label{sec:related}

This work connects two literatures that rarely cite each other, specification mining
and distributed runtime verification. It also draws on negative controls in statistics,
monitoring under imperfect observation, and the cost of false alarms.

\subsection{Specification mining}
Inferring properties from traces is a mature line of work. Daikon infers likely data
invariants at program points~\cite{daikon}. Perracotta mines two-event temporal patterns
from imperfect traces~\cite{perracotta}. Javert composes small patterns into larger
ones~\cite{javert}. Synoptic infers models constrained by invariants from
logs~\cite{synoptic}, and Texada mines arbitrary LTL templates~\cite{texada}. For timed
behaviour, \miner\ learns rules with bounded intervals~\cite{psiminer}. We adopt its
template in \eqref{eq:template}. This work mines centrally and offline and judges rules
as predictions about a dataset. Section~\ref{sec:distmine} treats mining over independent
observers with a communication budget.

\subsection{Ranking metrics and negative controls}
Coverage, confidence, support and lift come from association-rule
learning~\cite{apriori}. Its tendency to confuse co-occurrence with implication was
criticised early~\cite{correlations}. That critique is the origin of our shuffle test.
Destroying one structure while keeping others is also standard in time-series analysis,
where it is called surrogate-data testing~\cite{surrogate}. Temporal precedence together
with association is Granger's notion~\cite{granger}. It is weaker than
intervention~\cite{pearl}. Mined rules therefore support no causal reading. Sinha et
al.~\cite{csr} introduced the joint-row and independent-column controls and used them as
diagnostics. We keep that diagnostic role for the shuffle gap. Prop.~\ref{thm:inflate}
shows that it is a control on a fixed grid and not a physical end point. Compilation gets
its own criterion (Thm.~\ref{thm:gate}).

\subsection{Distributed and decentralised runtime verification}
Monitoring a property whose witnesses are spread across nodes is well
studied~\cite{sen2004,bauerfalcone,colombofalcone,themis}. For first-order properties,
MFOTL monitoring has been scaled offline~\cite{scalableoffline} and extended with
aggregation~\cite{monpolyagg}. BDD-encoded valuations make past-time first-order
monitoring efficient over unbounded domains~\cite{dejavu}. These works take a property as
given and ask how to monitor it across nodes. We ask which properties should exist,
whether each needs state, and which guarantees each needs. Hierarchical monitoring architectures for edge-IoT~\cite{hybrid,hierarchical} organise where such checks run. The closest work is
failure-aware distributed monitoring~\cite{failureaware}. Our concern complements it.
A transport that only \emph{disorders} evidence changes verdicts even when nothing fails.

\subsection{Monitorability and imperfect observation}
Whether violations can be detected from finite prefixes is the classical monitorability
question~\cite{pnuelizaks,rvltl,whatcanyouverify,adventures}. Another line accepts
imperfect observation. It infers missing states
probabilistically~\cite{rvse,adaptiverv,particlerv}, schedules monitoring under a
budget~\cite{samplingrv}, or handles recording uncertainty and lossy
traces~\cite{recordinguncertainty,lossytraces}. Thm.~\ref{thm:loss} belongs to this line
with a different question. Those works ask how to \emph{recover} a verdict despite missing
observations. We ask how a rule's own witness structure, its split into irreplaceable and
interchangeable events, determines its sensitivity to loss. The answer depends on how
full the rule's windows are and not on the size of its syntax.

These works mostly study missing events. Disorder has been treated as a correctness
problem for monitoring \emph{algorithms}. Runtime verification over out-of-order streams
reconstructs the verdict that a correctly ordered trace would give~\cite{outoforder}. Our
question is different, and as far as we know it has not been asked before. We ask how
much a disordering transport \emph{changes} the observed score of a mined rule, and
whether this can be predicted before deployment. Our claim is the link to the shuffle
statistic and its limits. That disorder affects monitoring is already known.

\subsection{Aggregation artefacts and measurement methodology}
Pooling heterogeneous observations can invent an association. This is an old warning in
statistics~\cite{simpson}. Here it takes a temporal form, because observers restart their
clocks, and distributed measurement depends on clock synchronisation~\cite{ntp,ptp}.
Section~\ref{sec:distmine} measures the cost at $79.9$ points and shows that offsetting
the clocks removes it exactly. Datasets of this kind are routinely concatenated. The
effect is therefore likely to occur in practice. The practical consequence is false alarms. Their
cost in intrusion detection is governed by the base-rate fallacy, and it has long limited
the deployment of learned detectors~\cite{baserate,closedworld}. Section~\ref{sec:fp}
measures our effect in those terms.

\section{Conclusion}\label{sec:conclusion}

Connecting specification mining to distributed monitoring changes the expected failure
mode. Coverage-based mining of rules evaluated within one observer distributes exactly,
provided the traces are separated in time and the rule descriptions arrive. Central mining
over merged traces is the step that needs justification. On a dataset whose observers
share a clock origin, merging inflated one rule from $11.6\%$ to $91.6\%$, and separating
the clocks removed the inflation to within $0.000$ points. Rule validation becomes a
compiler pass with two statistics. Keeping them apart allows a rule whose window is worth
$78$ points, but whose shuffle gap cannot be distinguished from zero, to be compiled
correctly. Under independent loss, detection depends on redundancy inside windows and
not on rule length. Contiguous loss, as real brokers produce it, did far more damage than
independent loss at the same rate. Reordering has no cheap summary. The same-event version
fixes where large displacement ends, and the shuffle gap bounds neither moderate nor
adversarial displacement. In practice we recommend mining where the data is observed,
gating each rule before deployment, and choosing the ordering guarantee from the range of
displacement that the deployment actually admits. On the fabric we measured, loss was the
limiting factor and ordering was not.

\section*{Artifact Availability}
The drivers, the chain evaluator, the synthetic generators, the broker harness and every
reported result will be made publicly available. The datasets~\cite{james} and the
mining engine~\cite{psiminer} are third-party and public. Only
Sect.~\ref{sec:envelope} needs running brokers.

\begin{acks}
This work has received funding from the European Union's Digital Europe Programme
(CYBERGUARD, No.~101190251) and from its Horizon Europe research and innovation programme
through the Smart Networks and Services Joint Undertaking (XTRUST-6G, No.~101192749).

Generative AI tools were used to help draft, edit and condense parts of the text. The
authors designed the study, proved the results, ran the experiments, and checked,
revised and approved all content. They take full responsibility for it.
\end{acks}
\raggedbottom
\balance
\appendix
\section{Technical Proofs}\label{app:proofs}
This appendix proves the two propositions of Sect.~\ref{sec:degrade}. No measured value
in the paper depends on it.

\begin{proof}[Proof of Prop.~\ref{thm:inflate}]
At $r=0$ the trace is unchanged. Take two distinct events $i\ne j$. The difference
$t_i'-t_j'$ of their displaced timestamps has a density bounded by $M/r$, where $M$
depends only on the density bound of $Z$. Hence
$\Pr(|t_i'-t_j'|\le k)\le 2Mk/r$, which tends to $0$. There are finitely many pairs. By a
union bound, with probability tending to one no two distinct events remain within $k$ of
each other. Every remaining witness then has all of its buckets satisfied by one event,
and these are exactly the witnesses of $\varphi^{(0)}$.

Two remarks complete the proof. First, $\varphi^{(0)}$ is a conjunction of per-event
conditions. Its score does not change under displacement. The limit is therefore a
statement about the original trace. Second, under continuous $Z$, two distinct events
receive equal displaced timestamps with probability zero. The limit thus equals
$\mathrm{score}(\varphi^{(0)})$, up to timestamp ties already present in the original
trace. The last claim of the proposition is the definition \eqref{eq:wgain}.
\end{proof}

\begin{proof}[Proof of Prop.~\ref{prop:refute}]
Fix $k>0$. Build a trace of $N$ disjoint pairs. Pair $i$ has an antecedent event at time
$a_i$ and a target event at $g_i = a_i + \tfrac{3}{2}k$. The pairs are separated so that
$a_{i+1} - g_i > 2k$. Each event is displaced independently by
$\varepsilon \sim U[-r,r]$.

First, $\mathrm{score}_0=0$. The only antecedent within $2k$ before $g_i$ is $a_i$, at
distance $\tfrac{3}{2}k>k$. No target is covered.

Second, $\Delta=0$ exactly. The shuffle changes which of the $2N$ timestamps carry the
antecedent attribute and leaves the timestamps in place. Every gap between two timestamps
is either $\tfrac{3}{2}k$ or larger than $2k$, so no two timestamps lie within $k$. No
assignment of attributes covers any target. Hence $\mathrm{score}_\pi=0$ for every
$\pi$, and $\Delta=0$.

Third, $\mathrm{score}_r>0$ for suitable finite $r$. Pair $i$ is covered when
$\varepsilon_{a_i}-\varepsilon_{g_i}\in[\tfrac{1}{2}k,\tfrac{3}{2}k]$. The difference of
two independent $U[-r,r]$ variables has a triangular distribution on $[-2r,2r]$. This
event therefore has positive probability whenever $r>k/4$.

We obtain $\mathrm{score}_r>\mathrm{score}_0=\lim_{r\to\infty}\mathrm{score}_r$, so
$\mathrm{score}_r$ is not monotone in $r$. We also obtain
$|\mathrm{score}_r-\mathrm{score}_0|>0=|\Delta|$, so the shuffle gap does not bound the
effect of displacement.
\end{proof}

\end{document}